\documentclass{svproc}

\usepackage{graphicx}
\usepackage{url}
\usepackage{float}

\begin{document}
\mainmatter              
\title{AI-Driven Cybersecurity Threats: A Survey of Emerging Risks and Defensive Strategies}
\titlerunning{AI-Driven Cybersecurity Threats}  
%
\author{Sai Teja Erukude\inst{1} \and Viswa Chaitanya Marella\inst{2} \and Suhasnadh Reddy Veluru \inst{2}}
\authorrunning{Sai Teja Erukude et al.} 
%
%

\institute{Bharat Institute of Engineering and Technology, Hyderabad, Telangana, India\\
\email{erukude.saiteja@gmail.com}\\
\and
Vellore Institute of Technology, Vellore, Tamil Nadu, India\\
\email{viswachaitanyamarella@gmail.com} \\
\email{suhasnadhreddyveluru@gmail.com}
}

\maketitle

\begin{abstract}
Artificial Intelligence's dual-use nature is revolutionizing the cybersecurity landscape, introducing new threats across four main categories: deepfakes and synthetic media, adversarial AI attacks, automated malware, and AI-powered social engineering. This paper aims to analyze emerging risks, attack mechanisms, and defense shortcomings related to AI in cybersecurity. We introduce a comparative taxonomy connecting AI capabilities with threat modalities and defenses, review over 70 academic and industry references, and identify impactful opportunities for research, such as hybrid detection pipelines and benchmarking frameworks. The paper is structured thematically by threat type, with each section addressing technical context, real-world incidents, legal frameworks, and countermeasures. Our findings emphasize the urgency for explainable, interdisciplinary, and regulatory-compliant AI defense systems to maintain trust and security in digital ecosystems.

\keywords{AI-Driven Cybersecurity, Cybersecurity Threats, AI-Enhanced Scams, Legislative Responses}
\end{abstract}

\section{Introduction}
AI allows new vulnerabilities in cybersecurity. Criminals, adversaries, and malicious actors use AI and other technologies to circumvent detection systems and automate attacks with minimal human intervention. The dual-use nature of AI raises major risks in the areas of cybersecurity, privacy, and public trust from those who exploit AI as a technological advancement for malicious use.

This study will discuss the growing cybersecurity risks that AI is creating, including deepfakes, adversarial attacks on machine learning models, and automating malware generation. We will demonstrate the increasing emergence of AI-facilitated social engineering attacks, including phishing and fraud, as well as a risk we call data poisoning, where an adversary can manipulate training data to compromise AI systems. Through case studies and discussion about defensive strategies and regulations, we hope to illustrate the lack of guarantees on a future response and the necessity of enforcement in the legal and technological response they need to defend against future vulnerabilities and risks.

\subsection{Survey Methodology}

The survey evaluated over 70 academic, industrial, and regulatory publications from 2017 to 2025, including peer-reviewed journals, preprint repositories, cybersecurity advisories, and whitepapers, and validated media outlets. The main selection criteria included: relevance to the threats of AI systems, such as deepfakes, adversarial attacks, automated malware, and AI-enabled scams; real-world solutions include case studies, toolkits, and frameworks (2019-2025); authority and credibility of sources, including governmental agencies.

\subsection{Comparative Taxonomy of Threats and Solutions}

Table \ref{tab_taxonomy_ai_threats} summarizes primary AI-driven threats mapped to attack modalities and defense strategies and forms the backbone of the subsequent sections.

\begin{table*}[ht]
\caption{Taxonomy of AI-Driven Cybersecurity Threats and Corresponding Defensive Strategies}
\setlength{\tabcolsep}{6pt}
\centering
\scriptsize
\begin{tabular}{|p{2cm}|p{2.5cm}|p{3cm}|p{3cm}|}
\hline
\textbf{Threat Category} & \textbf{Attack Vectors / Tools} & \textbf{Real-World Incidents} & \textbf{Defensive Strategies} \\
\hline
\textbf{Deepfakes \& Synthetic Media} & GANs, Voice Cloning, Face Swaps, Text-to-Video & Political deepfakes in Canada, AI-generated scams with celebrity voices \cite{sumsub}, \cite{mcAfee}, \cite{tribune} & XAI frameworks (n-gram analysis), wavelet-based detection, human-in-the-loop review, regulation (Digital India Act) \\
\hline
\textbf{Adversarial AI Attacks} & FGSM, PGD, C\&W, CleverHans, ART Toolkit, Poisoning & Adversarial image classification errors, model evasion in CAVs \cite{viso}, \cite{Hiddenlayer}, \cite{panda} & Adversarial training, Defensive distillation, Gradient masking, Certified robustness \\
\hline
\textbf{Automated Malware Generation} & FraudGPT, WormGPT, Polymorphic Engines, Obfuscators & AIIMS Ransomware (2022), WannaCry, BlackMamba AI malware \cite{ioactive}, \cite{medianama}, \cite{indiatoday} & EDR/XDR systems, AI-based behavior monitoring, automated incident response, UEBA \\
\hline
\textbf{AI-Powered Phishing \& Scams} & LLM-generated phishing, Deepfake voices, Chatbots & Pig Butchering scams, CEO voice fraud via AI~\cite{wired}, \cite{unodc}, \cite{mattburgers} & Weighted linguistic pattern detection, biometric authentication, scam detection models \\
\hline
\textbf{Social Engineering Automation} & AI Chatbots, Synthetic Identities, Face + Voice Deepfakes & \$25M Hong Kong executive fraud via Zoom deepfake, AI-driven dating scams~\cite{wired} & User education, deception detection software, image/audio verification, digital literacy \\

\hline
\end{tabular}
\label{tab_taxonomy_ai_threats}
\end{table*}

\section{Deepfakes and Synthetic Media} 
\label{section_deepfakes}

\subsection{The Rise in Deepfake Proliferation}
The rapid growth in AI-generated content has caused obstacles in industries, with the number of deepfake incidents rising tenfold worldwide from the year before \cite{sumsub}. North America, followed by the Asia-Pacific and Europe, experienced increases of 1740\%, 1530\%, and 780\%, respectively, with identity fraud, mostly about ID cards, being involved in almost 75\% of cases. 20\% of Americans have fallen for scams utilizing AI-generated celebrity endorsements, rising to 33\% for 18-34 year-olds \cite{mcAfee}. 1 in 4 Canadians encountered fake political content in the lead-up to the April 2025 election, including deepfake videos that erroneously show Prime Minister Mark Carney endorsing scams \cite{tribune}. The estimated rate of such attacks has been pegged at happening every five minutes.

\subsection{Detection Challenges and Observations}

\begin{figure}[ht]
    \centering
    \includegraphics[scale=0.3]{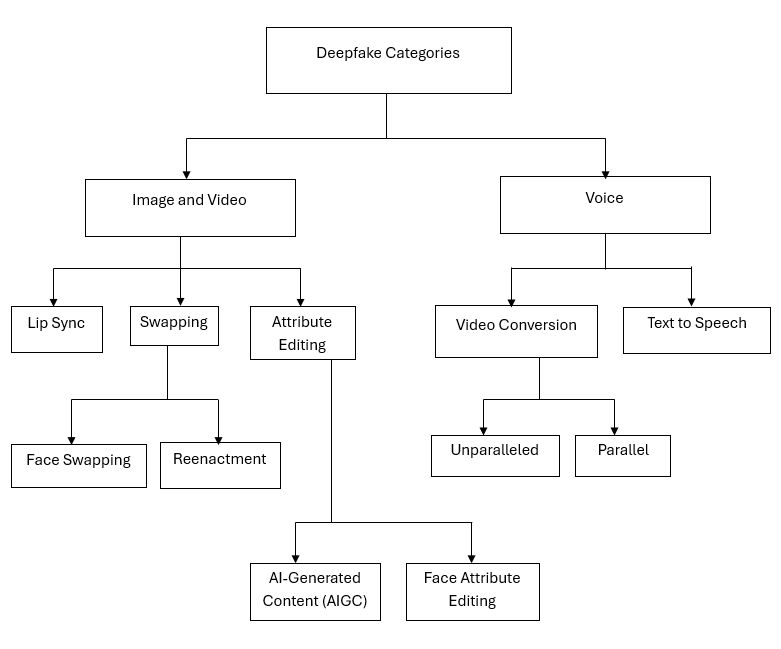}
    \caption{Different categories of deepfakes}
    \label{fig_deepfake_categories}
\end{figure}

Despite advancements in detecting deepfakes, current systems have serious limitations related to robustness and generalization. Benchmarks such as the ``Deepfake-Eval-2024" exhibit significant drops in the AUC score for video (50\%), audio (48\%), and image (45\%) in uncontrolled, real-world conditions \cite{chandra2025deepfakeeval2024multimodalinthewildbenchmark}. Many deep learning models rely on surface artifacts or background cues and thus are rendered ineffective when faced with sophisticated fakes. Wavelet-transformed feature extraction \cite{computers}, \cite{erukude2024}, shows promise to improve explainability and accuracy, but is still vulnerable to context, lighting, or language changes. Most deepfake detection systems are black box models and are subject to adversarial spoofing. Human-in-the-loop approaches have improved current detection capabilities, but the lack of a multi-modal approach and systematic enforcement of laws allows detection systems to be subject to zero-day manipulations and societal harm. As shown in Figure \ref{fig_deepfake_categories}, deepfakes encompass various forms, from manipulated video and audio to text and image-based fakes, highlighting the scope of challenges faced by detection frameworks

\subsection{Legal and Ethical Dimensions}

In the United States, only 14 states have passed anti-non-consensual sexual deepfake content laws, and only 10 have some restrictions on deepfakes relating to political campaigns. A bipartisan bill introduced in January 2024 aims to allow victims to sue the creators or distributors behind non-consensual sexual deepfake content, but with jurisdictional nuance that prevents legislated enforcement, due to ambiguity. India does not have specific deepfake laws, but some action can be taken under the Indian Penal Code and the Information Technology Act, and according to the Minister of State, AI-related regulations are coming under the new Digital India Act. South Korea has criminalized the distribution of so-called harmful deepfakes to the public interest under the premise of Article 245 of the Criminal Act, punishable by up to five years and/or a fine of 50 million won ($\approx$\$43,000). With evolving regulatory regimes and proposed laws, it is vital to accelerate public understanding of deepfake AI tools and to pursue change to combat the threat of deepfake fraud and manipulation.


\section{Adversarial AI Attacks}
\label{section_adversarial_attacks}

\subsection{Understanding Adversarial AI attacks}
Adversarial AI attacks refer to the perturbation of input data and methods that target weaknesses in AI models to produce incorrect or unwanted output. In this study, we argue that adversarial machine learning is the study of how adversarial examples (intentionally crafted input data) can successfully manipulate machine learning classifiers. When an adversary applies some perturbations to data at inference time to trick a trained model, the attack is called an evasion attack. In contrast, data poisoning is an adversarial attack where malicious samples are introduced into the training data to ``poison" the model and degrade its performance. This mechanism can be applied to any AI system and raises doubts regarding AI systems' reliability in safety-critical applications \cite{viso}.

\subsection{Attack Methods and Tools}

Adversaries have a variety of approaches for creating adversarial instances. For example, two common gradient-based approaches that provide undetectable perturbations to fool neural networks are the Fast Gradient Sign Method (FGSM) \cite{goodfellow} and Projected Gradient Descent (PGD) \cite{madry}. More advanced approaches like the Carlini \& Wagner (C\&W) attack can produce subtle and effective inputs that may go undetected \cite{carlini}. There are also multiple open-source frameworks that researchers and hackers have developed to assist with these attacks. For example, IBM's Adversarial Robustness Toolbox (ART), CleverHans \cite{papernot}, and Foolbox \cite{rauber} have dozens of attack algorithms in the image, text, and audio domains. The impact of adversarial manipulations is illustrated in Figure \ref{fig_adv_attack}, which visualizes how attackers perturb input data to deceive AI models.

\subsection{Defense Strategies}

Researchers have referred to the defense against adversarial attacks as an ``arms race" between the attacker and defender. Some key defense strategies are:

\begin{itemize}

\item \textbf{Adversarial Training:} Retraining models with adversarial examples seems to be the most developed and most successful defense strategy.

\item \textbf{Defensive Distillation:} Trains a secondary model on softened outputs to create smoother decision boundaries, increasing resistance to adversarial attacks, though adaptive threats may still bypass it \cite{neptune}.

\item \textbf{Adversarial Input Detection:} Identify and block adversarial inputs or escalate them for human review \cite{neptune}.

\begin{figure}[ht]
    \centering
    \includegraphics[scale=0.15]{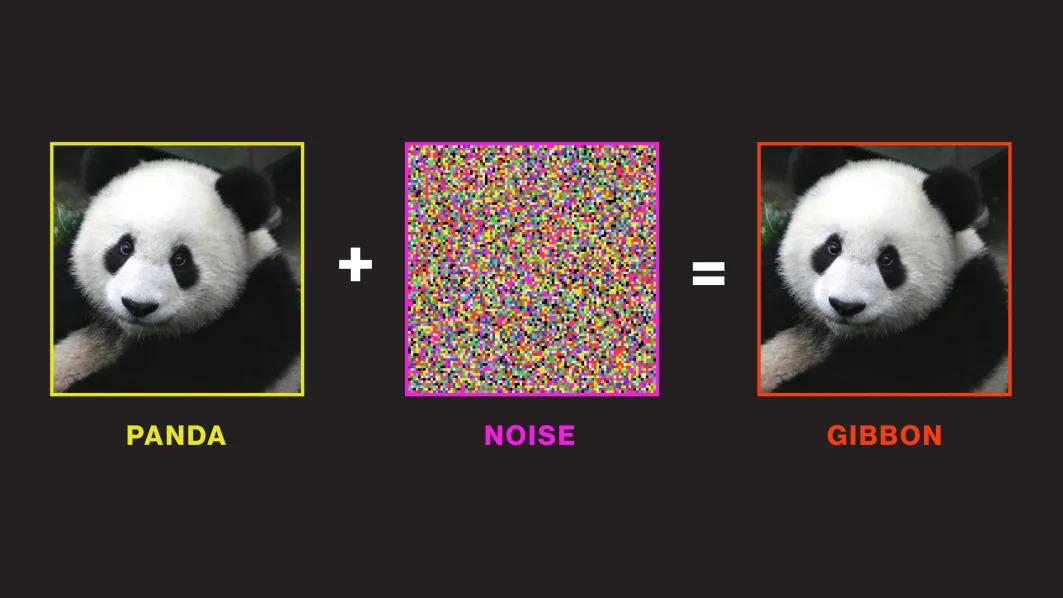}
    \caption{Adversarial attack manipulation}
    \label{fig_adv_attack}
\end{figure}

\end{itemize}

\subsection{Legal Perspectives: India and International}

India does not have any specific legislation governing cybersecurity risks from AI systems. The IT Act 2000 does not include emerging threats like data poisoning, and adversarial attacks will be prosecuted as a general cybercrime. The EU is ahead with its likely legislation; the draft AI Act indicates that high-risk AI systems have to prove they can withstand adversarial manipulation. The US continues to issue more guidelines than enforceable legislation, but the National Institute of Standards and Technology (NIST) is expected to release its adversarial threat defense recommendations in 2024. China is expected to propose limitations on the misuse of AI systems such as deepfakes. However, it is important to note that there have been no global laws governing adversarial AI.

\subsection{Critical Synthesis and Observations}

The space lacks a standardized approach to defending against attacks. Adversarial training is expensive and non-scalable. Defensive distillation or gradient masking have been proposed as low-cost, lightweight security measures, but they also have known vulnerabilities in adaptive attacks. The research community has developed a shared understanding that there are no ``silver bullets" and requires layered, context-specific security models. Points of divergence remain evident regarding the trade-offs among accuracy, interpretability, and robustness.


\section{Malware Generation and Automation}
\label{section_malware}

\begin{figure}[ht]
    \centering
    \includegraphics[width=6.5cm]{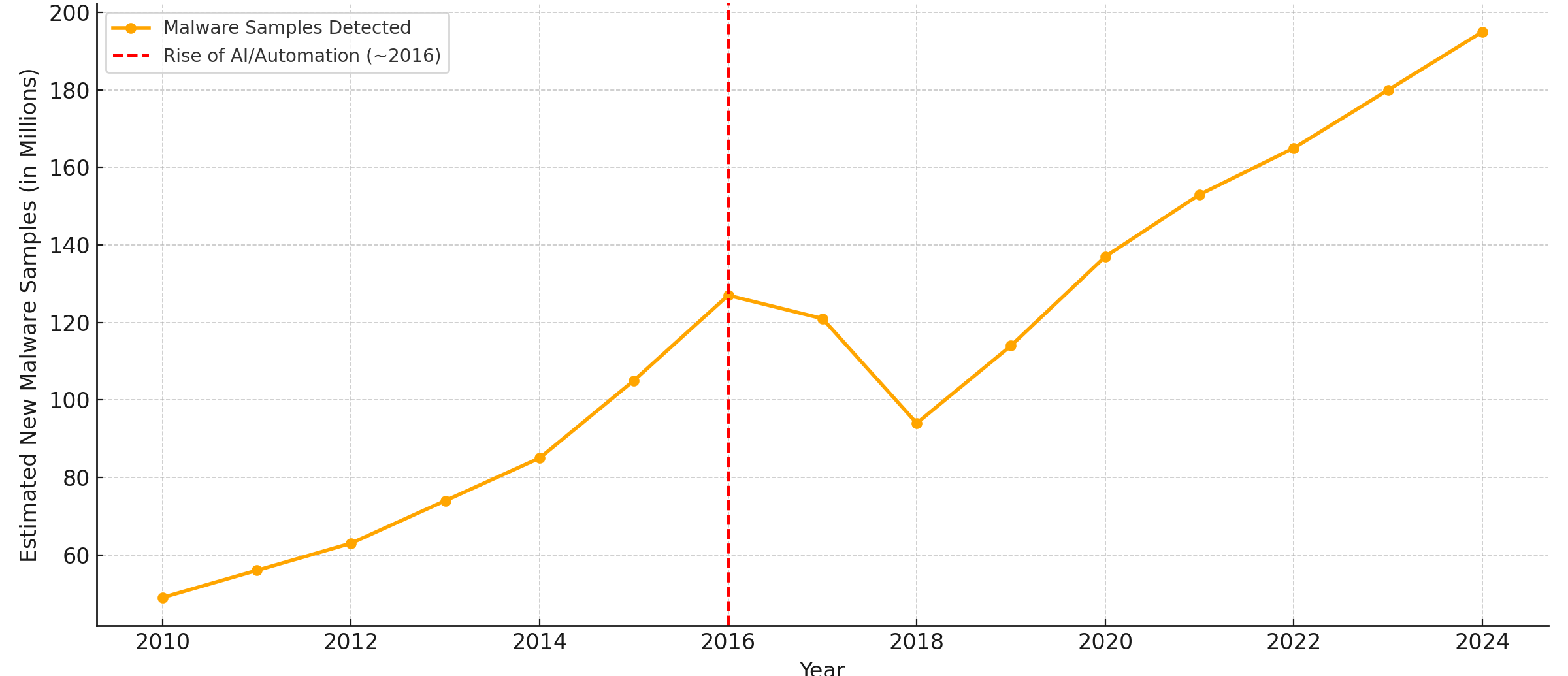}
    \caption{Global malware sample growth from 2010 to 2024}
    \label{fig_malware_growth}
\end{figure}

Modern malware commonly has polymorphic or metamorphic code, meaning it can change its structure without changing its underlying functionality, allowing the code to elude signature detection. New aspects of malware, such as obfuscation, and services like Malware-as-a-Service have made disseminating malware easier and detection harder. Generative AI extends this further and gives attackers the ability to generate or obfuscate malicious code automatically. Figure \ref{fig_malware_growth} illustrates that the difficulty posed by malware has become exponentially more complex since 2016, almost doubling \cite{AVTest}.

\subsection{Tools Available to Criminals for Malware Creation and Delivery}

\textbf{Crimeware frameworks:} Kits like Blackhole and Nuclear Pack \cite{darkreading} automate the creation and delivery of malware, and both have anti-detection and exploit toolkits \cite{hp}. Malware-as-a-Service sites make crime easier by creating a subscription-based model for malicious code and services.

\textbf{Botnets and automated delivery:} Botnets allow the delivery of malware and spam at a massive scale by taking advantage of networks of connected devices. The Mirai malware, for instance, infects IoT devices that use default credentials and creates a botnet to spread the malware. Other spam botnets like Emotet and TrickBot infect users by sending phishing emails.

\textbf{Obfuscation tools, and polymorphic engines:} Polymorphic malware changes its appearance every 30-60 seconds, producing a unique instance, thereby bypassing signature-based detection \cite{darkreading}, \cite{ioactive}. Table \ref{tab_obfuscation_tools} provides descriptions and common usages of widely deployed software obfuscation tools, highlighting their dual role in both legitimate software protection and malicious code evasion.

\begin{table}[ht]
    \caption{Descriptions and common usages of various software obfuscation tools}
    \setlength{\tabcolsep}{10pt} 
    \renewcommand{\arraystretch}{1.2} 
    \centering
    \scriptsize
    \begin{tabular}{|p{2cm}|p{4cm}|p{3cm}|}
    \hline
    \textbf{Tool Name} & \textbf{Description} & \textbf{Common Usage} \\
    \hline
    Themida & Commercial Protector combines packing and more. & DRM protection, malware obfuscation \\
    \hline
    VMProtect & Virtualizes and encrypts code execution flow. & High-end, DRM, software licensing \\
    \hline
    ConfuserEx & .NET protector with renaming, and resource encryption. & Malware, software IP protection \\
    \hline
    ProGuard & Java-based, shrinks, optimizes, and obfuscates bytecode & Android and app protection \\
    \hline
    JavaScript Obfuscator & Encrypts and inserts dummy JavaScript code. & Malicious web scripts, phishing campaigns \\
    \hline
    \end{tabular}
    \label{tab_obfuscation_tools}
\end{table}

\textbf{AI and ML–Driven Tools:} Tools like FraudGPT and WormGPT enable even those lacking technical skills to use AI to launch phishing campaigns and undetectable malware attacks.

\subsection{Recent Automated or Generated Malware Cases}

\begin{itemize}
\item \textbf{WannaCry Ransomware (2017):} WannaCry is one of the first large-scale malware outbreaks to include automated distribution, and leveraged a worm-based distribution process to propagate without human intervention. It infected more than 300,000 systems across more than 150 countries, with 48,000 in India alone, and has been estimated to have caused damages in the billions of dollars. The rapid, automated propagation of WannaCry used the EternalBlue vulnerability to overwhelm various defenses and make a timely and effective response and containment almost impossible \cite{indiatoday}.

\item \textbf{AIIMS Hospital Ransomware Attack (2022):} AIIMS Delhi was attacked with ransomware in November 2022 that encrypted 1.3 TB of data and impacted operations for multiple weeks. Due to ineffective network segregation, malware spreads across the network automatically. The October 2022 attack was likely ransomware-as-a-service and highlighted critical vulnerabilities in the cybersecurity posture of healthcare organisations \cite{medianama}.

\end{itemize}

\subsection{Defensive Technologies Against Automated Malware}

Organizations are increasingly adopting multi-dimensional, AI-enhanced defenses against rapidly evolving threats:

\begin{itemize}
\item \textbf{Antivirus and Heuristics:} AVs have introduced ML and behavioral heuristics for signature detection and to detect unmapped malware \cite{ioactive}.

\item \textbf{Endpoint Detection and Response (EDR):} EDR tools continuously observe behaviors on systems and identify anomalous behavior \cite{upwind}.

\item \textbf{Extended Detection and Response (XDR):} XDR examines data across endpoints and networks to identify complicated, multi-staged attacks.

\item \textbf{AI-Based Threat Detection:} ML can process massive amounts of telemetry and detect anomalies and zero-day threats in real-time.

\item \textbf{Behavioral Analytics:} UEBA can observe a user's and system's behavior, and identify anomalies that may indicate malware activity, even when there is no prior knowledge of the malware \cite{crowdstrike}.

\item \textbf{Automated Incident Response:} Playbooks are implemented to isolate compromised endpoints or to disable accounts and quickly contain threats.
\end{itemize}

\begin{figure}[ht]
    \centering
    \includegraphics[width=8.5cm]{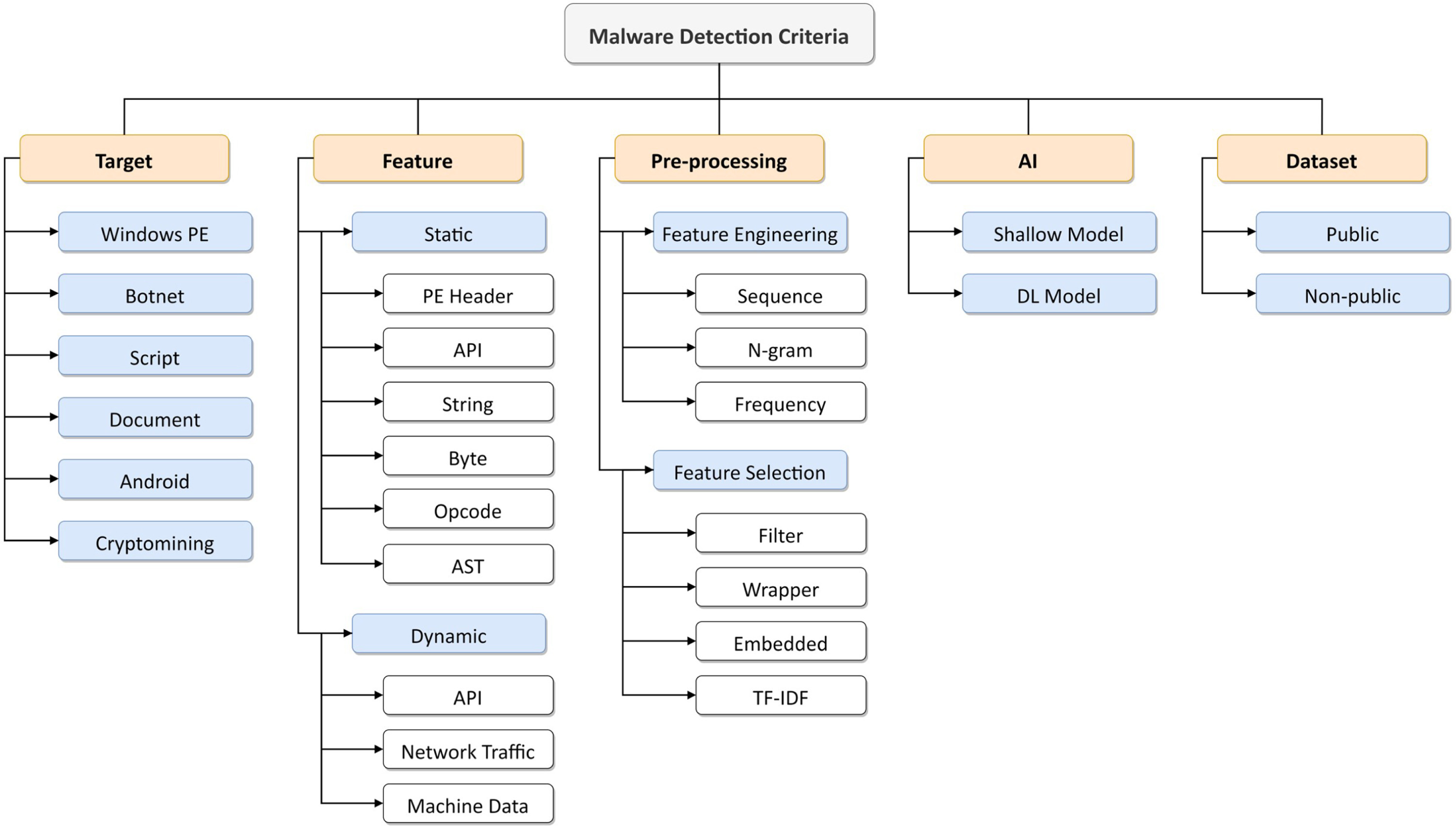}
    \caption{Malware Detection based on Target, Features, Pre-processing, AI, and Dataset.}
    \label{fig_malware_detection}
\end{figure}

The analytical pipeline of modern malware detection, including feature extraction, AI models, and dataset utilization, is summarized in Figure \ref{fig_malware_detection}.

\subsection{Legal and Law Enforcement Responses}

Cybersecurity law in India is a function of the IT Act, 2000, the Indian Penal Code (IPC), and the Digital Personal Data Protection Act (DPDPA), 2023 \cite{EY}. The DPDPA mandates data protection and penalties for breaches. The CERT-In Guidelines, 2022, require incident reporting within 6 hours for critical sectors \cite{clearias}. However, these frameworks do not address AI-generated malware (WormGPT, FraudGPT) or the ethical challenges of dual-use AI systems. Addressing these gaps may include expanding the definitions of the IT Act to include AI misuse, holding developers accountable. India could also benefit from cybercrime-specialized courts, digital forensics training for law enforcement, and mandatory AI risk assessments for critical systems.

\subsection{Critical Synthesis and Observations}

Overall, research is advancing toward AI being an effective option for anomaly detection; however, there is still some division regarding explainable AI and trust. Also, an asymptotic dance has begun between obfuscation (adversarial) methods and disengagement (heuristic) systems. Meanwhile, regulatory frameworks like India’s DPDPA regarding managing the risk of models of renaissance AI, and CERT-In, mention avoiding risk referred to earlier, but do not offer an answer on how to deal with autonomous, and or self-modifying malware.


\section{Automation of AI-powered Scams}
\label{section_scams}

\subsection{Phishing attacks}

With advancements in text-to-speech diffusion models, an attacker can clone a target voice in less than 30 seconds of source audio. The synthesized voice is then conveyed over VoIP or conferencing services, exploiting ``authority bias" while also bypassing traditional email-centric phishing filters. Recent high-profile incidents involving AI-driven voice cloning are outlined in Table \ref{tab:voiceclone_cases}, illustrating the significant financial and operational impacts across sectors.

\begin{table}[htbp]
    \centering
    \caption{Representative voice-cloning–enabled social-engineering incidents (2019–2025)}
    \label{tab:voiceclone_cases}
    \begin{tabular}{|c|c|p{2cm}|p{3.7cm}|p{3.7cm}|}
        \hline
        \textbf{Year} & \textbf{Loss} & \textbf{Victim} & \textbf{Exploit Vector} & \textbf{Short Take-away} \\
        \hline
        2019 & US\$243\,k & UK energy subsidiary & Deepfake audio of German CEO pressured CFO to expedite supplier payment & First widely reported AI-voice BEC; payment reached Mexico via Hungary shell account \cite{trendmicro}\\
        \hline
        2020 & US\$35\,m & UAE bank (HK branch manager) & Cloned director's voice plus forged e-mails convinced manager to authorise 17 wire transfers & Multi-modal spoofing (audio + e-mail) scaled to eight-figure fraud \cite{incidentdatabase}\\
        \hline
        2024 & US\$25\,m & Multinational (Hong Kong) & Full-body video deepfakes of CFO and colleagues in Zoom meeting & Liveness cues can be faked; triggered regional regulatory review \cite{incode}\\
        \hline
        2025 & $\le$US\$20\,m & Retail crypto investors & YouTube ``live" streams of deep-fake Elon Musk promising double-your-money giveaways & Convergence of celebrity deepfakes and mass-phishing; $>$15 verified channels detected \cite{cloudsek}\\
        \hline
    \end{tabular}
\end{table}

Explainable AI frameworks, such as the weighted n-gram analysis \cite{fi17040135}, can be adapted to detect linguistic patterns. By analyzing the similarities between known phishing attempts and unseen ones, this framework provides insights to identify and mitigate phishing campaigns.

\subsection{Personalized Social Engineering Attacks}

AI is changing social engineering by allowing bad actors to deploy personalized fraud attacks with frightening specificity. In a case more recent to Hong Kong, criminals deployed AI to commit fraud, involving deepfake video calls to impersonate executives and even romantic partners \cite{wired}. Victims described attending Zoom-style meetings where the entity on the other side appeared and blinked and smiled and spoke in familiar voices; all of this was fake to gain trust, encourage them to invest in the fake business opportunity, or provide sensitive data. One poor victim lost more than \$25 million when they thought they were attending a virtual meeting with the CFO of their own company.

\subsection{Pig Butchering Scams}

Pig butchering scams are long-term frauds where scammers develop levels of trust, often while pretending to be a romantic partner or financial advisor, before persuading them to deposit money, usually into cryptocurrency exchanges. The term pig butchering refers to ``fattening up" the victim emotionally before ``slaughtering" them financially. In 2024, scammers engaged multiple victims at the same time with AI chatbots, ensuring the emotional tone and language fluency stayed the same through consistency \cite{unodc}. The use of AI-generated images as well as video messages increased the scammers' credibility. One particular case detailed a U.S. tech executive defrauding him of \$1.2 million by using a realistic, AI-generated crypto platform using fake market data and customer assistance, which was gone when he tried to withdraw money.

\subsection{Automation}

The real danger of AI in cybercrime is that it's now possible to automate complex and large-scale scams. With AI, criminals can operate at a level of realism that enhances the social engineering experience for victims. For example, in a U.S.-documented pig butchering scam, an AI chatbot called ``Evelyn" had a months-long romantic conversation with a target while sending automated AI-edited selfies of itself and pre-recorded videos, all of which were for a fake crypto exchange \cite{mattburgers}. AI has reduced human involvement, enabling scalable fraud.

\section{Future Work}

Several promising directions for future research have emerged from this survey that require collaboration with AI researchers, cybersecurity practitioners, and government officials to create adaptable environments for our digital systems.

\begin{itemize}
    \item \textbf{Multi-Modal Explainable Detection Pipelines:} Develop architectures that can classify video, audio, text, and image data and balance robustness and interpretability.

    \item \textbf{Adaptive Threat Simulation and Benchmarking Platforms:} Create a standardized testing configuration across the public data to evaluate security models across diverse modalities and simulate zero-day threats.

    \item \textbf{Cross-Jurisdictional AI Risk Governance:} Support technical and legal communities in advancing regulatory frameworks across jurisdictions to bridge the disconnect between rapid AI innovation and policy landscapes.
    
\end{itemize}

\section{Conclusion}

AI has introduced a new era of cybersecurity challenges by morphing the threats into layered, intelligent, and scalable attack vectors. This paper has examined four types of AI-driven threats: deepfakes, adversarial attacks, automated malware, AI phishing, and AI social engineering. Collectively, we observe the evolving potential of adversarial actors to use generative, personalized, and autonomous capabilities to exploit traditional security-based infrastructure. Table \ref{tab:summary_recommendations} provides a summary of key recommendations for the threats discussed.

\begin{table}[H]
\caption{Summary of Threats, Gaps, and Suggested Interventions}
\setlength{\tabcolsep}{6pt}
\renewcommand{\arraystretch}{1.2}
\scriptsize
\centering
\begin{tabular}{|p{2.5cm}|p{3.8cm}|p{4.5cm}|}
\hline
\textbf{Threat Category} & \textbf{Key Gaps} & \textbf{Recommended Responses} \\
\hline
Deepfakes & Weak generalization, legal enforcement gaps & Human-in-loop detection, public awareness, AI content watermarking, regulatory acceleration \\
\hline
Adversarial AI & No universal defense, expensive adversarial training & Layered security, XAI, context-aware models, resilient model certification \\
\hline
Malware Automation & Obfuscation bypasses signature detection & Real-time AI anomaly monitoring, developer accountability \\
\hline
Phishing \& Scams & Hyper-personalized attacks, low user vigilance & Weighted n-gram filtering, continuous user education, voice biometrics \\
\hline
Social Engineering & Cross-modal deception, emotional manipulation & Real-time verification, scam simulation training, AI reporting portals \\
\hline
\end{tabular}
\label{tab:summary_recommendations}
\end{table}


\end{document}